\documentstyle[12pt,aps]{revtex}
\newcommand{\be}{\begin{equation}}
\newcommand{\ee}{\end{equation}}
\newcommand{\ba}{\begin{eqnarray}}
\newcommand{\ea}{\end{eqnarray}}
\newcommand{\bi}[1]{\bibitem{#1}}
\newcommand{\fr}[2]{\frac{#1}{#2}}
\newcommand{\non}{\nonumber}
\newcommand{\ar}{\mbox{$\rightarrow$}}
\def\vec#1{{\mbox{\boldmath$#1$}}}

\newcommand{\al}{\mbox{$\alpha$}}

\newcommand{\om}{\mbox{$\omega$}}
\newcommand{\Z}{\mbox{$Z\alpha$}}

\newcommand{\rp}{\mbox{$\vec{r}'$}}
\newcommand{\k}{\mbox{$\vec{k}$}}

\newcommand{\r}{\mbox{$\vec{r}$}}

\newcommand{\dE}{\mbox{$\Delta E$}}
\newcommand{\K}{\mbox{$\cal{K}$}}
\newcommand{\G}{\mbox{$\cal{G}$}}

\newcommand{\lb}{\left (}
\newcommand{\rb}{\right )}
\newcommand{\la}{\left\langle}
\newcommand{\ra}{\right\rangle}

\newcommand{\sS}{\mbox{$\vec{\sigma}\vec{\sigma}'$}}
\newcommand{\ga}{\vec{\gamma}}
\newcommand{\Ga}{\vec{\Gamma}}
\newcommand{\lm}{\mbox{$\lambda$}}
\newcommand{\E}{\mbox{$\cal{E}$}}

\begin{document}
\pagestyle{empty}
\hfill {\large BudkerINP-97-80}
\vspace{1.0cm}

\begin{center}
{\Large \bf Vacuum Polarization Contribution to Hydrogen and Positronium
Energies}

\bigskip

{\bf A.S. Yelkhovsky}\\
Budker Institute of Nuclear Physics,\\
and \\
Physics Department, Novosibirsk University, \\
630090 Novosibirsk, Russia
\end{center}

\bigskip

\begin{abstract}
Relative order $\al(\Z)^3$ shift of the energy levels induced by the vacuum
polarization is reexamined for a bound system of two particles with masses
$m$ and $M$. Recent results for hydrogen and for positronium are shown to
contain an error due to the inadequate procedure of the infrared divergence
handling. Numerically, the correction to the ground state energy constitutes
0.647 kHz for hydrogen and 46.7 kHz for positronium.
\end{abstract}

\bigskip


\bigskip

\newpage

\pagestyle{plain}
\pagenumbering{arabic}

\begin{center}
{\bf I. INTRODUCTION}
\end{center}

The radiative correction of the relative order $\al(\Z)^3$ to the energies of
positronium and the radiative-recoil correction of the relative order
$\al(\Z)^3\fr{m}{M}$ to the hydrogen energies should be taken into account
when one compares the recent experimental results \cite{posexp,hydexp} with
the QED predictions. An attempt to calculate such the corrections, induced
by the vacuum polarization, was undertaken in Ref.\cite{P} for hydrogen. In
Ref.\cite{EG5}, the same correction was calculated for a two-body system of
particles with an arbitrary mass ratio and the result was applied to
hydrogen and positronium.

In the present note the calculation of the vacuum polarization correction
is reexamined with the purpose to clarify the issue of the infrared
regularization. Such a regularization is necessary for a control over
the linear infrared divergence arising in the course of the calculation
and through which the leading order correction reveals itself.

In order to separate the contribution of the leading order, which comes
from the atomic scale, from the next-to-leading one, saturated by the
relativistic scale, we introduce the auxiliary parameter $\lm'$, $m\Z \ll
\lm' \ll m$. Subtracting the Yukawa potential from the Coulomb one,
\be
-\fr{\Z}{r}\;\;\; \ar\;\;\; -\fr{\Z}{r} + \fr{\Z}{r}e^{-\lambda' r},
\ee
we find the contribution to the energy due to the range of $r$'s satisfying
$\fr{1}{\lambda'} < r \sim \fr{1}{mZ\alpha}$. On the other hand, the
subtracted contribution is saturated by the short range,
$\fr{1}{\lambda'} > r \sim \fr{1}{m}$. In the sum of two contributions, the
dependence on the parameter $\lm'$ is cancelled away.

Below we use a standard scheme of calculations, taking the vacuum
polarization into account by means of the substitution
\be\label{subst}
\fr{g_{\mu\nu}}{k_0^2-\k^2}\;\;\; \ar \;\;\;
\fr{g_{\mu\nu}}{k_0^2-\k^2-\lm^2},
\ee
for a propagator of the photon, which polarizes vacuum. An energy correction
which thus becomes a function of \lm\ is then integrated over \lm\ with a
density of intermediate states $\rho(\lm)$, which equals
\be\label{rho}
\fr{2\al}{3\pi}\fr{\lm^2+2m^2}{\lm^3}\sqrt{1-\fr{4m^2}{\lm^2}}\theta(\lm-2m)
\ee
for a free particle-antiparticle pair.

\begin{center}
{\bf II. ORDER $m\alpha(Z\alpha)^4$ CONTRIBUTION}
\end{center}

If bounded particles are non-relativistic, vacuum polarization manifests
itself primarily due to the Coulomb field. Since the atomic momentum is much
less than a mass \lm\ of the particle-antiparticle pair, this momentum can
be neglected in the right-hand side of (\ref{subst}), so that an effective
potential induced by the Coulomb vacuum polarization, at fixed \lm\ turns
out to be
\be\label{V0}
V_{\lambda}(\r)=-\fr{4\pi\Z}{\lm^2} \delta(\r),
\ee
while the corresponding (lowest-order) contribution to an atom's energy is

\be
\E_{LO}(\lm) = -\fr{4\pi\Z|\psi(0)|^2}{\lm^2}.
\ee
After the integration with respect to \lm\ with the weight function
(\ref{rho}) it turns into
\be
\dE_{LO}=-\fr{4\al\Z|\psi(0)|^2}{15m^2}.
\ee

Now let us find the correction to this result induced by the modification of
the Coulomb potential at short ranges (recall that $\lm'\gg m\Z$),
\be\label{Cmod}
-\fr{\Z}{r}\;\;\; \ar\;\;\; -\fr{\Z}{r} + \fr{\Z}{r}e^{-\lambda' r}.
\ee
To this end we calculate the average value of the operator (\ref{V0}) over
the state, whose wave function is perturbed by the short-range correction to
the Coulomb potential (\ref{Cmod}):
\be
\E_{\lambda'} = \la \psi(\rp)\left|V_{\lambda}(\rp)\G\lb\rp,\r|E\rb
            \fr{\Z e^{-\lambda'r}}{r}\right| \psi(\r) \ra + c.c.
\ee
Here $\G\lb\rp,\r|E\rb$ is the reduced Green's function of the Schr\"odinger
equation in the Coulomb field, $\psi(\r)$ is the solution of this equation.
To the lowest order in \Z, we can disregard the Coulomb interaction in $\G$
and the atomic momentum as compared with $\lm'$:
\be
\G\lb 0,\r|E\rb \ar - \fr{2\mu}{4\pi r}, \;\;\;\; \psi(\r) \ar \psi(0),
\ee
($\mu=mM/(m+M)$ is the reduced mass) and obtain
\be\label{comp}
\E_{\lambda'} = \fr{4\pi(\Z)^2}{\lm^2}|\psi(0)|^2 \fr{4\mu}{\lm'}.
\ee

\begin{center}
{\bf II. ORDER $m\alpha(Z\alpha)^5$ CONTRIBUTION}
\end{center}

In the next-to-leading order, let us consider the double photon exchange
accounting for vacuum polarization by one of these photons.
According to the Feynman's rules, corresponding contribution to the energy
at fixed \lm\ is
\ba\label{skel}
\E (\lm,\lm')&=&-2(\Z)^2\psi^2\int\fr{dk_0}{2\pi i}\int\fr{d^3\k}{(2\pi)^3}
                \fr{4\pi}{k^2-\K'^2}\fr{4\pi}{k^2-\K^2}\la\gamma_{\mu}
                \fr{(1+\gamma_0)m+\gamma_0 k_0 -
                \ga\k}{k^2-\om^2}\gamma_{\nu}\ra         \non
                \\
        && \la\Gamma_{\mu}\fr{(1+\Gamma_0)M-
                \Gamma_0 k_0+\Ga\k}{k^2-\Omega_-^2}
                \Gamma_{\nu} - \lb \mu\leftrightarrow\nu, M\ar -M \rb \ra.
\ea
This expression can be represented graphically as two fermion lines
connected by two photon ones. The first and the second terms in the angle
brackets correspond to the graphs with uncrossed and crossed photon
propagators, respectively. In (\ref{skel}), $k^2 = \k^2$,
\[
\K = \sqrt{k_0^2-\lm^2}, \;\;\;\;\;\K' = \sqrt{k_0^2-\lm'^2};
\]
\[
\om = \sqrt{k_0^2+2mk_0}, \;\;\;\;\;
\Omega_{\pm} = \sqrt{k_0^2\pm 2Mk_0};
\]
$\gamma_{\mu}$ $(\Gamma_{\mu})$ are the Dirac matices for the light (heavy)
particle. The parameter \lm\ will be used below as a mass of the virtual
pair,
in accord with (\ref{subst}), while the parameter $\lm'$ is introduced to
regularize the otherwise infrared divergent integral in (\ref{skel}). Two
possible ways to insert \lm\ and $\lm'$ into the photon propagators are
accounted for in (\ref{skel}) by the overall factor 2.  Choosing
$\lm,\lm'\gg m\Z$, we can neglect atomic momenta so that taking the average
over a bound state reduces to that over the Pauli spinors which is denoted
by angle brackets, together with the multiplication by $\psi^2\equiv
|\psi(0)|^2$. The spinor averages are trivial:
\ba
&& \la\gamma_{\mu}(1+\gamma_0)\gamma_{\nu}\ra
   \la\Gamma_{\mu(\nu)}(1+\Gamma_0)\Gamma_{\nu(\mu)}\ra = 4, \\
&& \la\gamma_{\mu}\gamma_0\gamma_{\nu}\ra
   \la\Gamma_{\mu(\nu)}(1+\Gamma_0)\Gamma_{\nu(\mu)}\ra =
   \la\gamma_{\mu}(1+\gamma_0)\gamma_{\nu}\ra
   \la\Gamma_{\mu(\nu)}\Gamma_0\Gamma_{\nu(\mu)}\ra = 2, \\
&& \la\gamma_{\mu}\gamma_0\gamma_{\nu}\ra
   \la\Gamma_{\mu(\nu)}\Gamma_0\Gamma_{\nu(\mu)}\ra = 4 \mp 2\sS, \\
&& \la\gamma_{\mu}\ga\gamma_{\nu}\ra
   \la\Gamma_{\mu(\nu)}\Ga\Gamma_{\nu(\mu)}\ra = - 6 \pm 4\sS.
\ea

Before proceeding further, let us consider the analytic properties of the
integrand in (\ref{skel}) as a function of $k_0$. The photon propagators
have the poles at the points $\pm\sqrt{k^2+\lm^2}$ and
$\pm\sqrt{k^2+\lm'^2}$.
After the integration over $\k$ these poles turn into the cuts that go from
$-\infty$ to $-\lm$ $(-\lm')$ and from $\lm$ $(\lm')$ to $\infty$.
Similarly, the light fermion propagator gives rise to the cuts $(-\infty,
-2m]$ and $[0,\infty)$. The heavy fermion propagator in the second term of
(\ref{skel}), which corresponds to the graph with crossed photon lines, gives
rise to the cuts $(-\infty, -2M]$ and $[0,\infty)$. Finally, the heavy
fermion propagator in the first term, corresponding to the graph with
uncrossed photon lines, produces the cuts $(-\infty,0]$ and $[2M,\infty)$.

It is convenient to extract the heavy fermion propagator from the first term
of (\ref{skel}),
\be
\fr{1}{k^2-\om^2}\fr{1}{k^2-\Omega_-^2}=\fr{1}{2(M+m)k_0}\lb\fr{1}{k^2-\om^2}
- \fr{1}{k^2-\Omega_-^2}\rb,
\ee
and then to change the sign of the integration variable $k_0 \ar -k_0$ in
all terms containing $1/(k^2-\Omega_-^2)$. In this way the integral over
$k_0$ is naturally splitted into two parts. The former one,
\ba\label{ci}
&&\E_{cut}(\lm,\lm') = -2\fr{(\Z)^2\psi^2}{M^2-m^2}\int_{C_-}\fr{dk_0}{2\pi
i}
           \int\fr{d^3\k}{(2\pi)^3}\fr{4\pi}{k^2-\K'^2}\fr{4\pi}{k^2-\K^2}
            \\
         &&  \left[\lb\fr{2Mm}{k_0}+2m+\fr{m}{M}\fr{2k_0^2-k^2}{k_0}+
           \fr{\sS}{3}\fr{3k_0^2-2k^2}{k_0}\rb\fr{2M}{k^2-\Omega^2}
           - (M\leftrightarrow m) \right],  \non
\ea
where $\Omega\equiv\Omega_+$, is taken over the contour $C_-$, wrapping the
left cut. The latter part is a residue at the $k_0=0$ pole, which appears in
terms containing $k_0^{-1}(k^2-\Omega_-^2)^{-1}$ after the change $k_0 \ar
-k_0$:
\be\label{pi}
\E_{pole}(\lm,\lm')=-2\fr{(\Z)^2\psi^2}{M+m}\fr{4\pi}{\lm+\lm'}
                   \lb \fr{2Mm}{\lm\lm'}+1-\fr{2\sS}{3} \rb.
\ee

Let us begin with the latter contribution. In (\ref{pi}), the first term in
the brackets is singular when one of the lambdas approaches zero. This
term is just the regulator contribution subtracted from the leading order
correction. In fact, neglecting $\lm'$ as compared with \lm\ in the sum
$\lm+\lm'$ above, we see that the infrared--singular term in (\ref{pi}) is
compensated by the effect of the Coulomb potential modification
(\ref{comp}). Hence, only two last terms in the right-hand side of
Eq.(\ref{pi}) comprise the genuine order $(\Z)^5$ contribution of the
$k_0=0$ pole, so that we can safely set $\lm'=0$ in those terms:
\be\label{pimod}
\E_{pole}(\lm) = -2\fr{(\Z)^2\psi^2}{M+m}\fr{4\pi}{\lm}
                   \lb 1-\fr{2\sS}{3} \rb.
\ee

Likewise, we can set $\lm'=0$ in the left-cut contribution (\ref{ci}) as far
as this procedure does not spoil the infrared convergence of the integral.
The integration over \k\ gives for $\E_{cut}(\lm)\equiv\E_{cut}(\lm,0)$:
\ba
&&\E_{cut}(\lm)=\fr{4(\Z)^2\psi^2}{M^2-m^2}\int_{C_-}dk_0
                \left\{\lb\fr{2m}{k_0}+\fr{m}{M}\rb\fr{1}{|k_0|+\K}
                \right. \non \\
                &&\left. -\left[\fr{2Mm}{k_0^2}+\fr{m}{M}
                -\fr{\sS}{3}\lb\fr{4M}{k_0}-1\rb\right]\fr{1}{\Omega+\K}
                -(M\leftrightarrow m) \right\}.
\ea
Finally, integrating with respect to $k_0$ and adding up the pole
contribution (\ref{pimod}), we obtain:
\ba\label{fin}
\E(\lm)&=&-\fr{4(\Z)^2\psi^2}{M^2-m^2}\left\{Mm(M-m)\fr{2\pi}{\lm^3}-
          \fr{4Mm}{\lm^2}A\lb\fr{\lm}{2M}\rb+(M-m)\fr{2\pi}{\lm}\fr{\sS}{3}
          \right.\non \\
          &&+\fr{m}{M}\lb\fr{1}{2}+\ln\fr{\lm}{M}\rb+\sS\left[\ln\fr{M}{m}-
          \fr{4}{3}A\lb\fr{\lm}{2M}\rb\right]\non \\
          &&\left.-\lb\fr{m}{M}+\fr{\sS}{3}\rb
          \fr{\lm^2}{2M^2}\left[A\lb\fr{\lm}{2M}\rb+\ln\fr{\lm}{M}\right]
          -(M\leftrightarrow m) \right\}.
\ea
Here
\be
A(x)=\theta (1-x) \fr{\sqrt{1-x^2}}{x} \cos^{-1} x  -
      \theta (x-1) \fr{\sqrt{x^2-1}}{x} \cosh^{-1} x .
\ee

From (\ref{fin}), the known results for the vacuum polarization contribution
to the hyperfine splitting in muonium \cite{CL,TY} and positronium
\cite{STY} can be obtained:
\ba
\dE_{hfs}^{\mu^+ e^-}&=&\al(\Z) E_F \left\{ \fr{3}{4} - \fr{m}{M}
                        \lb 2\ln^2\fr{M}{m} +\fr{8}{3}\ln\fr{M}{m}
                        +\fr{28}{9} + \fr{\pi^2}{3} \rb
                        +{\cal O}\lb\fr{m^2}{M^2}\rb \right\};\\
\dE_{hfs}^{e^+ e^-}&=&\fr{5}{3}\al(\Z) E_F .
\ea
Here
\be
E_F=\fr{8\pi}{3}\fr{\Z\psi^2}{Mm}
\ee
is the Fermi splitting with the anomalous magnetic moments omitted.

For the spin-independent part of the correction to the energy levels of
hydrogen and positronium, we have
\ba
\dE^{p^+ e^-}&=&\fr{4(\Z)^2\psi^2}{Mm}\int_1^{\infty} dx \rho(x)\left\{
           \fr{\pi}{2x^3}-\fr{1}{x^2}+\fr{\sqrt{x^2-1}}{x^3}\cosh^{-1}x
           \right.\non \\
  &&\left.+\fr{1}{2}+\ln 2x-2x^2\ln 2x+2x\sqrt{x^2-1}\cosh^{-1}x\right\}\\
              &=& \fr{\al(\Z)^2}{\pi}\fr{\psi^2}{Mm}\lb\fr{47\pi^2}{144}
                  -\fr{70}{27} +{\cal O}\lb\fr{m}{M}\rb\rb;\label{hyd}
\ea
and
\ba
\dE^{e^+ e^-}&=&-\fr{2(\Z)^2\psi^2}{m^2}\int_1^{\infty} dx \rho(x)\left\{
              \fr{\pi}{2x^3}-\fr{\cosh^{-1}x}{x^3\sqrt{x^2-1}}-\fr{1}{x^2}-2
              \right.\non \\
              &&\left.-2\ln 2x+2x^2\ln 2x-8x\sqrt{x^2-1}\cosh^{-1}x
                -2\fr{x\cosh^{-1}x}{\sqrt{x^2-1}}\right\}\\
              &=&\fr{\al(\Z)^2}{\pi}\fr{\psi^2}{m^2}\lb\fr{49\pi^2}{288}
                  -\fr{40}{27} \rb,\label{pos}
\ea
where
\[
\rho(x)=\fr{\al}{3\pi}\fr{2x^2+1}{x^4}\sqrt{x^2-1}.
\]
These results differ from those obtained in Refs. \cite{P,EG5} for hydrogen
and in Ref. \cite{EG5} for positronium. The error made in both works has the
same origin -- inaccurate treatment of the infrared divergence. In fact, the
authors of Refs. \cite{P,EG5} do not introduce in (\ref{skel}) the parameter
$\lm'$, which, as we have seen above, regularizes the infrared divergence.
Instead, they subtract from the integrand in (\ref{skel}) its asymptotic
value at small $k$'s. Giving the finite results of Refs. \cite{P,EG5}, this
last procedure cannot be correct, since those finite results arise as a
difference between two divergent integrals. In contrast, the regularization
procedure used in the present work deals with the well-defined finite
expressions.

Numerically, the correction (\ref{hyd}) constitutes 0.647 kHz for the
ground-state energy of hydrogen, while (\ref{pos}) equals 46.7 kHz for the
ground-state energy of positronium. In the case of hydrogen, the correction
exceeds the uncertainty of the recent measurement \cite{hydexp}.

\bigskip

\begin{center}
{\bf ACKNOWLEDGMENTS}
\end{center}

Partial support from the Russian Foundation for Basic Research (Grant No.
97-02-18450), and from the Universities of Russia Program (Grant No.
95-0-5.5-130) is gratefully acknowledged.

\end{document}